\documentclass[12pt]{article}
\usepackage{epsfig}
\usepackage{latexsym}
\begin{document}
\title{Solitons in the modified Yang-Mills theory.}
\author{ A.A.Slavnov \thanks{E-mail:$~~$ slavnov@mi.ras.ru}
\\Steklov Mathematical Institute, Russian Academy
of Sciences\\ Gubkina st.8, GSP-1,119991, Moscow,\\
 Moscow State University\\
 and Theoretical Division of CERN} \maketitle

\begin{abstract}
It is shown that pure Yang-Mills theory in the modified formulation
admits soliton solutions of classical field equations.
\end{abstract}

\section {Introduction}

It is well known that the theory of the Yang-Mills field noninteracting
with other fields does not produce soliton excitations (\cite{Col}, \cite{Des}, \cite{Pag}).
On the other hand lattice simulations indicate that confinement of color is related
to the existence of solitons.

Recently the new formulation of non-Abelian gauge theories applicable beyond perturbation
theory and allowing the unique quantization procedure free of Gribov ambiguity (\cite{Gr}, \cite{Si})
was proposed (\cite{Sl1}, \cite{Sl2}, \cite{QS}). The usual arguments forbidding the existence
of solitons in the Yang-Mills theory are not applicable to this formulation.

In this paper we are going to show that the solitons of the t'Hooft-Polyakov (\cite{H}, \cite{P}) magnetic
monopole type indeed are possible in this formulation.

The paper is organised as follows. In the second section the new formulation of non-Abelian
gauge theories is presented, and its application to the Yang-Mills theory in the framework
of perturbation theory is discussed. In the third section we demonstrate the existence of
topologically nontrivial soliton solutions of classical field equations in the model under
consideration. Concluding remarks are given in the Discussion.

\section{The model.}

We start with the non-Abelian model, described by the Lagrangian
\begin{equation}L=-\frac{1}{4}F_{\mu\nu}^i F_{\mu\nu}^i+\frac{1}{2}D_\mu \varphi^i D_\mu \varphi^i
-\frac{1}{2}D_\mu \chi^i D_\mu \chi^i+iD_\mu b^i D_\mu e^i
\label{1}
\end{equation}
 For simplicity we consider the case of $SU(2)$ gauge group. Here
$F_{\mu\nu}$ is the standard curvature tensor for the Yang-Mills
field. The scalar fields $\varphi,\chi, b, e$ belong to the adjoint
representation of the group $SU(2)$. The fields $\varphi, \chi$ are
commuting, and the fields $b, e$ are anticommuting. In this equation
$D_{\mu}$ is the usual covariant derivative, hence the Lagrangian
(\ref{1}) is gauge invariant. Note the minus sign before the terms,
describing the field $\chi$. Because of the sign they possess  the
negative energy.

We assume (and check later) that the fields, entering the Lagrangian (\ref{1}) have the following asymptotic behaviour
\begin{equation}
|\varphi|\rightarrow|\frac{m}{g}|; \quad
|\chi|\rightarrow|\frac{m\alpha}{g}|; \quad r=|\textbf{x}|; \quad
r\rightarrow \infty \label{2}
\end{equation}
The parameter $\alpha\rightarrow 1$ when $g\rightarrow 0$ sufficiently fast. For example we may take
\begin{equation}
\alpha=\frac{g^{-n}-g^n}{g^{-n}+g^n}=1-g^{2n}+ \ldots \label{3}
\end{equation}
So that $1-\alpha= O(g^{2n})$, and choosing $n$ big enough we get in
a formal perturbation theory the results, coinciding with the
standard Yang-Mills theory to arbitrary order in $g$. In the
eq.(\ref{2}) $m$ is a constant having dimension of mass.

Speaking about formal perturbation theory, we assume formal series
in the coupling constant, no matter is it convergent or not. This is
the usual notion for quantum field theory. If the coupling constant
$g$ is small, like in quantum electrodynamics it means that the
usual relations between the elements of the scattering matrix
(unitarity, causality e.t.c.) are approximately fulfilled in a given
order of formal perturbation theory. But in the theories like
quantum chromodynamics (QCD) the coupling constant is not small and
even the separate terms in the formal perturbation series may not
exist due to infrared divergencies. Nevertheless one usually insists
on the formal relations like unitarity and causality in QCD. This
point of view is supported by the fact that correlators of the gauge
invariant operators as a rule have no infrared singularities. One
may hope that for a proper choice of asymptotic states these
problems may be avoided.

In a topologically trivial sector corresponding to the perturbation
theory we can choose the direction, where the asymptotic does not
vanish, as the third axis in the charge space. Making the shifts of
the fields preserving manifest Lorentz invariance
\begin{equation}
\varphi^i= \tilde{\varphi^i}+\delta^{i3}mg^{-1}; \quad \chi^i=\tilde{\chi^i}-\delta^{i3}m\alpha g^{-1}
\label{4}
\end{equation}
we get the Lagrangian with fields $\tilde{\varphi}, \tilde{\chi}$
going to zero at $r\rightarrow \infty$. This is necessary to develop
a perturbation expansion near the vacuum state.

We want to prove that the scattering matrix obtained after this
shift in the framework of perturbation theory coincides with the
usual scattering matrix in the Yang-Mills theory. If $\alpha \neq
1$, we may speak about the coincidence of the scattering matrices up
to arbitrary order in formal perturbation theory. For any term in
the perturbative expansion of the scattering matrix we may choose
$n$ big enough to get the complete agreement up to this term. In the
Yang-Mills theory the scattering matrix does not exist due to
infrared divergencies, but one can nevertheless speak about the
absence of transitions between the states including only physical
excitations and the states including some unphysical ones. Physical
excitations in both theories are three dimensionally transversal
components of the Yang-Mills field.

Instead one can consider the correlation functions of gauge invariant operators which do not suffer from infrared singularities.
We shall show that they also coincide in both theories up to arbitrary order in perturbation theory.

At the same time in topologically nontrivial sectors these theories differ: the Yang-Mills theory in the standard formulation
has no  soliton excitations, but the modified theory has classical solitons.

For $\alpha \neq 1$ the theory we consider is not the standard
Yang-Mills theory. But it is gauge invariant for any $\alpha$, and
the values of observables calculated using a formal perturbation
theory in the coupling constant coincide to any order in the
coupling constant with the values, calculated in the usual
Yang-Mills theory. Moreover this formulation may be used beyond
perturbation theory and does not suffer from the Gribov ambiguity
\cite{Sl2}. If the coupling constant $g$ is very small, and the
limit at $\alpha \rightarrow 1$ exists for observables, as it
happens in the electro-weak models based on the Higgs-Brout-Englert
mechanism  (\cite{Hi},\cite{BE}), no solitons are required. But in
quantum chromodynamics this limit does not exist for the on shell
scattering matrix elements due to infrared divergencies, and the
coupling constant $g$ is not small, so that $\alpha$ may differ
considerably from unity. In this case as we shall show soliton
excitations may arise.

The Lagrangian describing the modified theory after the shift (\ref{4}) looks as follows
\begin{eqnarray}
 L=-\frac{1}{4}F_{\mu\nu}^iF_{\mu\nu}^i+ D_\mu \tilde{\varphi}_+^iD_\mu \tilde{\varphi}_-^i+iD_\mu \tilde{b}^i D_\mu \tilde{e}^i+\nonumber\\
  +m\frac{1+\alpha}{\sqrt{2}}D_\mu \tilde{\varphi}_+^i\varepsilon^{ij3}A_\mu^j+m\frac{1-\alpha}{\sqrt{2}}D_\mu \tilde{\varphi}_-^i\varepsilon^{ij3}A_\mu^j
  + \frac{m^2(1-\alpha^2)}{2}A^a_\mu A^a_\mu
\label{5}
\end{eqnarray}
where obvious notations
\begin{equation}
\tilde{\varphi}_\pm^i=\frac{\tilde{\varphi^i} \pm \tilde{\chi^i}}{\sqrt{2}}
\label{6}
\end{equation}
were introduced. In this equation $i,j=1,2,3$ and $a=1,2$.

This Lagrangian for any $\alpha$ is invariant with respect to " shifted" gauge transformations
\begin{eqnarray}
\delta A_\mu^i=\partial_\mu \eta^i+g\varepsilon^{ijk}A_\mu^j \eta^k \nonumber\\
\delta \tilde{\varphi}_-^1=-\frac{1+\alpha}{\sqrt{2}}m\eta^2+g\varepsilon^{1jk}\tilde{\varphi}_-^j\eta^k \nonumber\\
\delta \tilde{\varphi}_+^1=-\frac{1-\alpha}{\sqrt{2}}m\eta^2+g\varepsilon^{1jk}\tilde{\varphi}_+^j\eta^k \nonumber\\
\delta \tilde{\varphi}_-^2=\frac{1+\alpha}{\sqrt{2}}m\eta^1+g\varepsilon^{2jk}\tilde{\varphi}_-^j\eta^k \nonumber\\
\delta \tilde{\varphi}_+^2=\frac{1-\alpha}{\sqrt{2}}m\eta^1+g\varepsilon^{2jk}\tilde{\varphi}_+^j\eta^k \nonumber\\
\delta \tilde{\varphi}_-^3=g\varepsilon^{3jk}\tilde{\varphi}_-^j\eta^k \nonumber\\
\delta \tilde{\varphi}_+^3=g\varepsilon^{3jk}\tilde{\varphi}_+^j\eta^k \nonumber\\
\delta \tilde{b}^i=g\varepsilon^{ijk}\tilde{b}^j\eta^k \nonumber\\
\delta \tilde{e}^i=g\varepsilon^{ijk}\tilde{e}^j\eta^k
\label{7}
\end{eqnarray}

In perturbation theory there is no Gribov ambiguity, so we can
choose the gauge $\partial_\mu A^i_\mu=0$ introducing also the
Faddeev-Popov ghosts $\bar{c}^i, c^i$.

The scattering matrix for $\alpha=1$ may be presented by the path integral
\begin{eqnarray}
 S=\int d\mu\{i[\int d^4x (-\frac{1}{4}F_{\mu\nu}^iF_{\mu\nu}^i+ D_\mu \tilde{\varphi}_+^iD_\mu \tilde{\varphi}_-^i+\nonumber\\
+\lambda^i\partial_\mu A_\mu^i+i\partial_\mu\bar{c}^i D_\mu
c^i+iD_\mu \tilde{b}^iD_\mu \tilde{e}^i +m\sqrt{2}D_\mu
\tilde{\varphi}_+^i\varepsilon^{ij3}A_\mu^j)]\} \label{9}
\end{eqnarray}
where the measure $d\mu$ is the product of differentials of all the fields.

Expanding the scattering matrix for  arbitrary $\alpha$ over $1-\alpha$ near the point $1=\alpha$ one has
\begin{eqnarray}
S=\int d\mu \exp\{i\int
dx[[L^{ef}_{\alpha=1}][1+\nonumber\\(1-\alpha)[m^2\int dx
(A_\mu^i)^2+\int dx\frac{m}{\sqrt{2}}D_\mu
\tilde{\varphi}_-^i\varepsilon^{ij3}A_\mu^j]+ \ldots]\} \label{9a}
\end{eqnarray}
where the effective action is given by the eq.(\ref{9}). We ignore the ultraviolet divergencies,
having in mind that some invariant ultraviolet regularization (dimensional or higher derivative) is introduced.

Therefore the second term and the following terms have at least the order $1-\alpha=O(g^{2n})$.
So we conclude that the results obtained with the help of Lagrangian (\ref{5}) coincide with the usual ones
at least to the order $2n$. As the number $n$ is arbitrary that means the formal perturbation expansion obtained in this way coincides with the usual one.

Being interested in the perturbative results we may put $\alpha=1$, as $\alpha=1-O(g^{2n})$, where $n$ is an arbitrary number.
Clearly in this case no mass term for the Yang-Mills field is generated, as due to the opposite signs of the terms depending
on $\varphi$ and $\chi$ their contributions to the mass of the Yang-Mills field cancel.

 For $\alpha=1$ the action is also invariant with respect to supersymmetry transformation
\begin{equation}
 \delta \tilde{\varphi}_-^i=i\tilde{b}^i\epsilon; \quad \delta\tilde{e}^i=\tilde{\varphi}_+^i\epsilon;
 \quad \delta\tilde{b}^i=\delta \tilde{\varphi}_+^i=0.
 \label{8}
 \end{equation}
It is easy to see that these transformations are nilpotent
\begin{equation}
\delta^2\tilde{\varphi}_-^i=0; \quad \delta^2\tilde{e}^i=0
\label{8a}
\end{equation}
This invariance provides the decoupling of excitations corresponding to the fields $\tilde{\varphi}_\pm, \tilde{b}, \tilde{e}$.

According to the Noether theorem these symmetries generate conserved
charges $Q_B, Q_S$. The corresponding asymptotic conserved charges
are denoted as $Q_B^0$ and $Q_S^0$ and the asymptotic states may be
chosen to satisfy the equations
\begin{equation}
Q_B^0|\psi>_{ph}=0; \quad Q_S^0|\psi>_{ph}=0; \quad
[Q^0_B,Q^0_S]_+=0 \label{12}
\end{equation}
where $Q_B^0$ and $Q_S^0$ are the asymptotic charges equal to
\begin{equation}
Q_B^0=\int
d^3x[(\partial_iA_0-\partial_0A_i)^j\partial_ic^j-\lambda^j\partial_0c^j]
\label{13}
\end{equation}
\begin{equation}
Q_S^0=\int d^3x(\partial_0\varphi_+^ib^i-\partial_0b^i\varphi_+^i)
\label{14}
\end{equation}
 The second
equation (\ref{12}) provides the decoupling of the excitations
corresponding to the fields $\varphi_\pm, e, b$.
 The first equality is analogous to the corresponding equality in the BRST treatment of
the standard Yang-Mills theory. It guarantees the absence of the
transitions from the states containing only the excitations
corresponding to the transversal components of the Yang-Mills field
 to the states containing longitudinal and temporal quanta.

The explicit  form of the asymptotic conserved charges may be
obtained as follows.  For matrix elements, which do not include
excitations, corresponding to the fields $\varphi_{\pm}, b, e$ the
last term in the eq.(\ref{9}) does not contribute into the conserved
charges. That means for any states which do not include the
excitations, corresponding to the fields $\tilde{\varphi_\pm^i}$ and
$e,b$ the matrix elements of the scattering matrix may be calculated
with the Lagrangian in the exponent (\ref{9})without the last term.

The explicite expression for asymptotic charges may be obtained as
follows. Rescaling the fields
\begin{equation}
\tilde{\varphi}_+^i=\varphi_+^i(m\sqrt{2})^{-1}; \quad
\tilde{\varphi}_-^i=\varphi_-^i (m\sqrt{2}); \quad
\tilde{b}^i=b^i(m\sqrt{2}); \quad \tilde{e}^i=e^i (m\sqrt{2})^{-1}
\label{10}
\end{equation}
we have in the exponent the following effective Lagrangian
\begin{eqnarray}
L=-\frac{1}{4}F_{\mu\nu}^iF_{\mu\nu}^i+ D_\mu \varphi_+^iD_\mu \varphi_-^i+\nonumber\\
 +\lambda^i\partial_\mu A^i_\mu+i\partial_\mu\bar{c}^iD_\mu c^i+ +iD_\mu b^iD_\mu e^i +D_\mu \tilde{\varphi}_+^i\varepsilon^{ij3}A_\mu^j
\label{11}
\end{eqnarray}
After this change the Lagrangian (\ref{11}) does not depend on $m$.
Boundary conditions for the states, which do not contain the quanta of the fields $\varphi$ and $b,e$
also do not change after such transformation. That means the integral (\ref{9}) does not depend on $m$
and we can put $m=0$. Putting in the eq.(\ref{9}) $m=0$, we have in the exponent the action,
which is invariant with respect to the usual BRST-transformations and the super transformations (\ref{8}).

Any vector, satisfying eqs. (\ref{12}- \ref{14}) has a structure
\begin{equation}
|\psi>_{ph}=|\psi>_{tr}+|N>
\label{15}
\end{equation}
where $|\psi>_{tr}$ is a vector which contains only transversal quanta and $|N>$ is a zero norm vector.
Factorising this subspace with respect to the vectors $|N>$, we get the physical space which coincide
with the space of states of the Yang-Mills theory. So we proved the perturbative unitarity of the Yang-Mills
scattering matrix in the space, which contains only physical excitations.
The proof however was formal, as the on shell scattering matrix elements do not exist due to infrared
singularities. We can however speak about nullification of the matrix elements corresponding to transitions
 between physical and unphysical states.

The only nontrivial sensible objects in the perturbative Yang-Mills theory are correlation functions of gauge invariant operators.
One can easily see that these correlation functions coincide in the standard and modified formulations up to arbitrary order in perturbation theory.
Indeed, one can repeat the consideration given above and see that  these correlation functions  are given by the path integrals
\begin{eqnarray}
 Z=\int d\mu\{\exp[i\int dx(-\frac{1}{4}F_{\mu\nu}^iF_{\mu\nu}^i+ D_\mu \varphi_+^iD_\mu \varphi_-^i+\nonumber\\
 +\lambda^i\partial_\mu A^i_\mu+i\partial_\mu\bar{c}^iD_\mu c^i+ +iD_\mu b^iD_\mu e^i+J(x)O(x))]\}
\label{16}
\end{eqnarray}
where $O(x)$ is some gauge invariant function, depending only on $A_\mu(x)$, and $J(x)$ is a source.
Boundary conditions for all fields in eq.(\ref{16}) correspond to the vacuum states.

In the generating functional (\ref{16}) we can perform the integration over $\varphi_\pm, e, b$. The terms, which arise after such integration cancel,
as the fields $\varphi_\pm$ and $b,e$ have an opposite statistics, and we get the standard expression for the generating functional of the correlation
functions of gauge invariant operators.
\begin{equation}
Z=\int d\tilde{\mu}\{\exp[i\int
dx(-\frac{1}{4}F_{\mu\nu}^iF_{\mu\nu}^i +\lambda^i\partial_\mu
A^i_\mu+i\partial_\mu\bar{c}^iD_\mu c^i+O(x)J(x))]\} \label{16a}
\end{equation}
where the measure $d\tilde{\mu}$ includes product of differentials
 \begin{equation}
 d\tilde{\mu}=dA_\mu^id\lambda^jd\bar{c}^kdc^m
 \label{16b}
 \end{equation}
 The same conclusion may be obtained if we work in the gauge, applicable beyond perturbation theory, for example
$\varphi_-^a=0, a=1,2; \quad \partial_\mu A_\mu^3=0$, so even beyond
perturbation theory one can pass freely to any admissible gauge, for
example to the gauge $A_0=0$.

\section{Classical solitons in the topologically nontrivial sector.}

In this section we show that the model considered above produces nontrivial soliton excitations of the t'Hooft-Polyakov
magnetic monopole type.

We consider the classical action corresponding to the Lagrangian
(\ref{1}) and look for the classical solitons with the asymptotic
for large $r$
\begin{equation}
\varphi^i\rightarrow\frac{x^im}{rg}; \quad \chi^i\rightarrow-\frac{x^im\alpha}{rg}
\label{17}
\end{equation}
We are working with the stationary solutions in the gauge $A_0=0$.

Contrary to the previous section, where formal perturbation expansion was used, we shall work in the topologically nontrivial sector and consider the soliton solutions of classical equations of motion
\begin{eqnarray}
D_jF^l_{ij}+g\varepsilon^{ijm}(D_j\varphi)^l \varphi^m-g\varepsilon^{ijn}(D_j \chi)^l\chi^n=0;
 \quad A^l_i\rightarrow \varepsilon^{lik}\frac{x^k}{gr^2}, r\rightarrow \infty \nonumber\\
D_i(D_i \varphi)^n=0; \quad \varphi^j(x)\rightarrow\frac{x^jm}{gr}, r\rightarrow \infty\nonumber\\
D_i(D_i\chi)^n=0; \quad \chi^j(x)\rightarrow -\frac{\alpha x^jm}{gr}, r\rightarrow \infty.
\label{18}
\end{eqnarray}
These boundary conditions provide decreasing of covariant
derivatives of the fields $\varphi, \chi$, which is important for
the finitness of the energy.To guarantee the finiteness of the
energy we consider only solutions which are non singular at
$r\rightarrow 0$. Now we cannot neglect the terms which are small in
a formal perturbation expansion as we are looking for solutions
which cannot be obtained in perturbation theory. We also note that
in practice this construction is applied to quantum chromodynamics,
where the coupling constant $g$ is not small.

We  shall  use the t'Hooft-Polyakov ansatz
\begin{eqnarray}
A^i_j(x)=\varepsilon^{ijk}\frac{x^k}{r}W(r); \quad \varphi^i(x)=\delta^{ji}\frac{x_j}{r}F(r) \nonumber\\
\chi^i(x)=\delta^{ji}\frac{x_j}{r}G(r); \quad A^i_0(x)=0, \nonumber\\
r\rightarrow \infty, W(r)\rightarrow (gr)^{-1}, F(r)\rightarrow F\cosh{\gamma}, G(r)\rightarrow F\sinh{\gamma},\nonumber\\ F\cosh{\gamma}=\frac{m}{g}; \quad F\sinh{\gamma}=-\alpha \frac{m}{g}.
\label{19}
\end{eqnarray}
 If $g$ is small, $\alpha\rightarrow 1$, as in the
electro-weak models based on the Higgs-Brout-Englert mechanism
(\cite{Hi}, \cite{BE}, \cite{QS1},) then $\varphi(x)\simeq \chi(x)$
and the equation for the Yang-Mills field has the same form as in
the standard theory. This equation has no soliton solutions. It
agrees with the statement that these theories are well described by
the perturbation series and do not require the existence of
solitons.

However we may also consider the theories, where $g$ is not small (as QCD).
Now the parameter $\alpha$ is different from 1, and our previous arguments are not valid. In these cases the parameters $F,\gamma$ are finite.

The equations (\ref{18}) may be rewritten in terms of the functions
\begin{equation}
K(r)=1-grW(r); \quad J(r)=F(r)rg; \quad Y(r)=G(r)rg
\label{20}
\end{equation}
\begin{eqnarray}
r^2\frac{d^2K}{dr^2}=(K^2+J^2-Y^2-1)K(r); \quad K(r)\rightarrow 0, r\rightarrow \infty\nonumber\\
r^2\frac{d^2J}{dr^2}=2K^2J; \quad J(r)\rightarrow Frg\cosh{\gamma};  r\rightarrow \infty \nonumber\\
r^2\frac{d^2Y}{dr^2}=2K^2Y; \quad Y(r)\rightarrow Frg\sinh{\gamma}=-\alpha Frg\cosh{\gamma};  r\rightarrow \infty
\label{21}
\end{eqnarray}
Following the paper \cite{JZ} we take the following ansatz  for the solutions
\begin{eqnarray}
J(r)=\Lambda(r)\cosh{\gamma}; \quad Y(r)=\Lambda(r)\sinh{\gamma};\nonumber\\
\Lambda(r)\cosh{\gamma}\rightarrow Frg\cosh{\gamma}; \quad \Lambda(r)\sinh{\gamma}\rightarrow Frg\sinh{\gamma}.
\label{22}
\end{eqnarray}

Thus the equations (\ref{21})acquire the form
\begin{eqnarray}
r^2\frac{d^2K}{dr^2}=(K^2+\Lambda^2-1)K; \quad K\rightarrow 0, r\rightarrow \infty, \nonumber\\
r^2\frac{d^2\Lambda}{dr^2}=2K^2\Lambda; \quad \Lambda(r)\rightarrow Frg; \quad r\rightarrow \infty.
\label{23}
\end{eqnarray}
The solutions of these equations are well known (\cite{PZ},\cite{H},\cite{P},\cite{Bog})
\begin{equation}
K(r)=\frac{rgF}{\sinh{rgF}}; \quad \Lambda(r)=\frac{rgF}{\tanh{grF}}-1.
\label{24}
\end{equation}

Obviously these solutions possess positive and limited energy, namely they have the same energy as the magnetic monopole
\begin{eqnarray}
E=\int d^3x[\frac{1}{4}F_{lm}^iF_{lm}^i+\frac{1}{2}(D_l\varphi)^i(D_l\varphi)^i -\frac{1}{2}(D_l\chi)^i(D_l\chi)^i]=\nonumber\\
\int d^3x[\frac{1}{4}F_{lm}^iF_{lm}^i+\frac{1}{2}(D_l\Lambda)^i(D_l\Lambda^i)]
\label{25}
\end{eqnarray}

Using the gauge invariant definition for electromagnetic field tensor \cite{H}
\begin{equation}
F_{\mu\nu}=\hat{\Lambda}^iF_{\mu\nu}^i-g^{-1}\varepsilon^{ijk}\hat{\Lambda}^i(D_\mu\hat{\Lambda})^j(D_\nu\hat{\Lambda})^k
\label{26}
\end{equation}
where $\hat{\Lambda}^i=\frac{\Lambda^i}{|\Lambda|}; \quad |\Lambda|=(\sum_i\Lambda^i\Lambda^i)^{1/2}$
we found the excitation we consider is the magnetic monopole, which produces the magnetic field
\begin{equation}
B^i(x)=\frac{x^i}{gr^3}
\label{27}
\end{equation}
One sees that even for $g$ large the mass and magnetic field of monopole solution do not depend on $\gamma$, and are
determined by the constants $F$ and $g$.

The solution (\ref{22}-\ref{24})has no electric charge. It is easy to include into this scheme also the excitations
possessing electric and magnetic charges, dyons\cite{JZ}.

\section{Discussion}

In this paper we showed that the modified formulation of the
Yang-Mills theory indeed admits soliton excitations. Let us remind
here our starting points. We are looking for the nonperturbative
soliton solutions which corresponds to the gauge invariant
Lagrangian generating the same perturbation series as the standard
Yang-Mills theory. The on-shell matrix elements of the scattering
matrix in perturbation theory formally coincide, but for QCD
perturbatively do not exist due to infrared divergencies. The
correlation functions of the gauge invariant operators are free of
this problem and coincide in the framework of perturbation theory in
the both theories. Modified formulation in the topologically
nontrivial sector has the classical solutions, corresponding to
solitons.

Of course our treatment of solitons was purely classical and there
are many questions to be answered. The main question is related to
the existence in our formulation of negative energies. In
perturbation theory we were able to show that the negative energy
states decouple from the positive energy ones.

Let us note that the same question arises in the standard Yang-Mills theory.
In renormalizable manifestly Lorentz invariant formulation the time-like quanta,
having the negative energy are present. In the framework of perturbation theory
we can prove their decoupling using for example the BRST quantization.
Beyond perturbation theory it is an open question. There is no problem to
introduce the indefinite metrics, which makes the eigenvalues of the Hamiltonian
positive, but the question about the existence of transitions
between different states is open.

 \textbf{Acknowledgements} This
work was completed, when the author was visiting the Theoretical
Division of CERN. I wish to thank for hospitality L. Alvarez-Gaume
and I.Antoniadis. Useful discussions with L.Faddeev, V.Rubakov  and
R.Stora are acknowledged. The work was supported in part by RFFI
grant 14-01-00695 A.
 \end{document}